\documentclass[preprintnumbers, floatfix, preprintnumbers, letterpaper, superscriptaddress,nofootinbib]{revtex4}
\usepackage{}
\usepackage{graphicx}
\usepackage{microtype}
\usepackage{amsmath}
\usepackage{amssymb}
\usepackage{subfigure}
\usepackage{hyperref}
\usepackage{url}
\usepackage{xcolor}
\usepackage{color}
\usepackage{mathrsfs}
\usepackage{calrsfs}
\usepackage{amsfonts}
\usepackage{eufrak}
\usepackage{eucal}
\usepackage{latexsym}
\usepackage{ragged2e}
\usepackage{epsfig}
\usepackage{textcomp}
\usepackage{makeidx}
\usepackage{array}
\usepackage{booktabs}
\usepackage{multirow}
\usepackage{threeparttable}
\usepackage{caption}
\DeclareCaptionJustification{justified}{\leftskip=0pt \rightskip=0pt \parfillskip=0pt plus 1fil}
\definecolor{vividviolet}{rgb}{0.62, 0.0, 1.0}
\definecolor{amaranth}{rgb}{0.9, 0.17, 0.31}
\definecolor{palatinateblue}{rgb}{0.15, 0.23, 0.89}
\definecolor{brightpink}{rgb}{1.0, 0.0, 0.5}
\definecolor{cornflowerblue}{rgb}{0.39, 0.58, 0.93}
\definecolor{deepcarminepink}{rgb}{0.94, 0.19, 0.22}
\definecolor{radicalred}{rgb}{1.0, 0.21, 0.37}

\hypersetup{linktoc=all,
    colorlinks, linkcolor={palatinateblue},
    citecolor={brightpink}, urlcolor={amaranth}
}

\graphicspath{{Images/}}

\graphicspath{{Images/}}

\def\@fnsymbol#1{\ensuremath{\ifcase#1\or \ddagger \or  $\textleaf$  \or \dagger
\else\@ctrerr\fi}}%

\makeatother

\def\sideremark#1{\ifvmode\leavevmode\fi\vadjust{\vbox to0pt{\vss
 \hbox to 0pt{\hskip\hsize\hskip1em
 \vbox{\hsize1.3cm\tiny\raggedright\pretolerance10000
 \noindent #1\hfill}\hss}\vbox to8pt{\vfil}\vss}}}%
\def\beq{\begin{equation}}
\def\eeq{\end{equation}}

\begin{document}

\title{The Hawking-Page-like Phase Transition from FRW Spacetime \\ \vspace{0.4cm} to McVittie Black Hole}

\author{Haximjan Abdusattar}
\affiliation{College of Science, Nanjing University of Aeronautics and Astronautics, Nanjing, 211106, China}

\author{Shi-Bei Kong}
\affiliation{College of Science, Nanjing University of Aeronautics and Astronautics, Nanjing, 211106, China}

\author{Yihao Yin}
\email{yinyihao@nuaa.edu.cn}
\affiliation{College of Science, Nanjing University of Aeronautics and Astronautics, Nanjing, 211106, China}

\author{Ya-Peng Hu}
\email{huyp@nuaa.edu.cn}
\affiliation{College of Science, Nanjing University of Aeronautics and Astronautics, Nanjing, 211106, China}
\affiliation{Key Laboratory of Aerospace Information Materials and Physics (NUAA), MIIT, Nanjing 211106, China}

\begin{abstract}

In this paper, we investigate the thermodynamics especially the Hawking-Page-like phase transition of the McVittie space-time. We formulate the first law of thermodynamics for the McVittie black hole, and find that the work density $W$ of the perfect fluid plays the role of the thermodynamic pressure, i.e. $P$:=$-W$. We also construct the thermodynamic equation of state for the McVittie black hole. Most importantly, by analysing the Gibbs free energy, we find that the Hawking-Page-like phase transition from FRW spacetime to McVittie black hole is possible in the case $P>0$.

\end{abstract}

\maketitle

\section{Introduction}

Black hole thermodynamics has been an active field of research for decades, but most of the earlier research focused only on those in asymptotic flat or (A)dS space. In this paper, we will investigate thermodynamics of black holes in a more realistic environment, that is, the FRW universe, which was not often seen in the literature but nevertheless important to do. The motivation is obvious -- for example, it has been conjectured that primordial black holes (PBHs) may be an important component of the dark matter (DM) \cite{Sasaki:2016jop,Niemeyer:1999ak}, and to understand how they were created in the early universe it is crucial to study a black hole in the context of cosmology that
(i) the actual universe is not empty, but instead full of matter field that interacts with the black hole; (ii) the radius of the black hole is affected by the change of the expansion rate of the universe over time \cite{Carr:1974nx}. In particular, we will study the thermodynamics of McVittie spacetime \cite{McVittie:1933zz}, which was the first solution to Einstein field equations that describes a Schwarzschild black hole embedded in a FRW universe filled with perfect fluid.

The McVittie spacetime usually contains a black hole horizon and a cosmological horizon, both of which are apparent horizons whose radii are time-dependent, and both can be studied from thermodynamic aspects.
The geometrical properties of McVittie spacetime have already been discussed extensively \cite{Nolan:1998,Kaloper:2010ec,daSilva:2012nh,Faraoni:2012gz,Faraoni:2015ula,Antoniou:2016obw,Piattella:2015xga,Gregoris:2019oxz,Ruiz:2020yye}, but on the contrary, its thermodynamic properties have rarely been studied. As far as we know, the unified first law of a dynamic and spherically symmetric spacetime was first proposed by Hayward \cite{Hayward:1997jp} and was later applied to McVittie spacetime \cite{Ke-Xia:2009tzo,Akbar:2017ljn}. Furthermore, its Hawking temperature was worked out in \cite{DiCriscienzo:2007pcr,Faraoni:2007gq}. However, many other important thermodynamic aspects have not yet been thoroughly investigated for McVittie spacetime, e.g. the thermodynamic equation of state and phase transitions of a black hole.

Besides the laws of thermodynamics, to describe the behavior of a thermodynamic system, it is usually crucial to construct its equation of state. This is often difficult for a black hole, unless we find a proper definition of its thermodynamic pressure. In recent research on asymptotically AdS black hole, the thermodynamic pressure has often been defined, with a lot of success, as a quantity proportional to the cosmological constant \cite{Kastor:2009wy,Dolan:2010ha,Cvetic:2010jb,Kubiznak:2012wp} (see also \cite{Altamirano:2014tva,Kubiznak:2016qmn} for reviews, and also \cite{Gunasekaran:2012dq,Wei:2012ui,Hendi:2012um,Altamirano:2013ane,Cai:2013qga,Bhattacharya:2017nru,Hu:2018qsy,Li:2020xkh} for related works). However, this kind of definition is not guaranteed to work for McVittie spacetime, since the cosmological constant can be set to zero in this context. With this difficulty in mind, in this paper we will extract the definition of the thermodynamic pressure from the first law, which turns out to be related with the work density of the perfect fluid. With such definition, we will further construct the thermodynamic equation of state for McVittie black hole.

After working out the thermodynamic equation of state of McVittie black hole, we will be able to further investigate its phase transitions. In particular, the main focus of this paper will be the Hawking-Page (HP) type of phase transition, which is the key to study the formation and evolution of a black hole. In \cite{Hawking:1982dh} Hawking and Page first proposed this interesting phase transition on AdS background between black holes and thermal radiation. According to their work, the black hole is only stable above a critical temperature. Below such a temperature the thermal AdS instead of the black hole becomes the stable one. This phase transition was later widely investigated in various black hole systems \cite{Cai:2007wz,Czinner:2017tjq,Mejrhit:2019oyi,Zhao:2020nrx,Li:2020khm,Spallucci:2013jja,Su:2019gby,Xu:2020ubo,Belhaj:2020mdr,Wei:2020kra}. In this paper, we will follow the same criterion as \cite{Spallucci:2013jja,Su:2019gby,Mbarek:2018bau,Marks:2021fpe} to determine the possibility of HP-like phase transition from FRW spacetime to McVittie black hole, that is, to determine their relative stability by comparing their Gibbs free energy with a fixed thermodynamic pressure. Our main conclusion is that such HP-like phase transition can happen if the thermodynamic pressure is positive.

The organization of this paper is as follows. In Sec.\ref{sII}, we will first briefly introduce the apparent horizons of McVittie spacetime. Then we focus on its black hole horizon and obtain the corresponding Hawking temperature and first law of thermodynamics.
In Sec.\ref{sIII}, we further derive the equation of state for the black hole, which shall be intensively investigated in Sec.\ref{sIV}, where the HP-like phase transition will be found and the stability of the black hole will be discussed. Sec.\ref{sVI} will be conclusions and discussions. In addition to the main text, the thermodynamics of the cosmological horizon will be discussed in Appendix \ref{A}.

\section{Review: Black Hole Horizon and Its First Law of Thermodynamics in McVittie Spacetime} \label{sII}

In this section, we will briefly review the apparent horizon, the Hawking temperature and the first law of thermodynamics in the McVittie spacetime.

We begin with the Einstein field equations
\begin{equation}\label{Eeq}
\mathcal {R}_{\mu\nu}-\frac{1}{2}g_{\mu\nu}\mathcal {R} =8\pi T_{\mu \nu} \,,
\end{equation}
where $\mathcal {R}_{\mu\nu}$ is the Ricci tensor, $\mathcal {R}$ is the Ricci scalar, $T_{\mu \nu}$ is the energy momentum tensor of the perfect fluid given by
\begin{equation}\label{Tmu}
T_{\mu \nu}=(\rho+p)u_{\mu}u_{\nu}+p g_{\mu \nu}\,,
\end{equation}
where $u^\mu=\frac{1}{\sqrt{-g_{tt}}}\delta_t^\mu$ is the four-velocity of the fluid, $\rho$ is its energy density, and $p$ is its pressure.

The McVittie spacetime is a spherically symmetric and time-dependent solution of the Einstein field equations, which describes a black hole embedded in a spatially flat FRW universe sourced by a perfect fluid.
In an isotropic coordinate system $x^\mu=$($t$, $r$, $\theta$, $\phi$), the line element of this spacetime is in the following form
\cite{McVittie:1933zz,Nolan:1998,Kaloper:2010ec,Nandra:2011ug}
\begin{equation}\label{MFRW}
d s^2= -\frac{\left[1-\frac{m}{2 a(t)r} \right]^2}{\left[1+\frac{m}{2 a(t)r}\right]^2}d t^2 +a^2(t) {\left[1+\frac{m}{2 a(t)r}\right]^4}[d r^2 +r^2 (d\theta^2 + \sin^2 \theta d\phi^2)] \,,
\end{equation}
where $m$ is the mass parameter which we assume to be a positive constant, and $a(t)$ is a time-dependent scale factor. Note that if $a(t)$ is constant, the McVittie spacetime reduces to the standard Schwarzschild black hole, or if $m\rightarrow0$, it reduces to the spatially flat FRW universe.
By introducing the areal radius
\begin{equation}\label{R}
R(t,r) \equiv a(t){\left[1+\frac{m}{2 a(t)r}\right]^2} r \,,
\end{equation}
the metric (\ref{MFRW}) can be rewritten as
\begin{equation}\label{nMFRW}
d s^2 =h_{i j}dx^{i}dx^{j}+R^2(d\theta^2 + \sin^2 \theta d\phi^2)\,,
\end{equation}
with
\begin{equation}
h_{i j}={\rm diag}\left[-\frac{(1-\frac{m}{2 a r})^2}{(1+\frac{m}{2 a r})^2}, a^2 \left(1+\frac{m}{2 a r}\right)^{4}\right],\label{hij}
\end{equation}
where $i,j=0,1, x^{0}=t, x^{1}=r, a=a(t), R=R(t,r)$.

From (\ref{Eeq}), (\ref{Tmu}) and (\ref{nMFRW}), one can derive the Friedmann's equations~\cite{Nolan:1998,Faraoni:2012gz,Kaloper:2010ec,Faraoni:2007es}
\begin{eqnarray}\label{GRT}
H^2=\frac{8\pi}{3}{\rho} \,,
\end{eqnarray}
and
\begin{eqnarray}\label{GRTp}
\dot{H}=-4\pi(\rho+p){\sqrt{1-\frac{2 m}{R}}} \,.
\end{eqnarray}
Note that here $p=p(t,R)$, and from (\ref{GRT}) we have $\rho=\rho(t)$, whose relation is given by \cite{Nandra:2011ug,Gregoris:2019oxz}
\begin{equation}\label{omega}
p(t,R)=\rho(t)\left[\frac{1+\omega(t)}{\sqrt{1-{2m}/R}}-1\right] ,
\end{equation}
where $\omega(t)$ is an equation of state parameter of the perfect fluid.

In order to study the thermodynamics of McVittie spacetime, we further need to find out its horizons, which serve as boundaries of thermodynamic systems. In McVittie spacetime,
the apparent horizon condition is given by \cite{Hayward:1994bu}
\begin{equation}
	h^{i j}\partial_{i}R\partial_{j}R=0 \,,\label{ahcond}
\end{equation}
and by substituting (\ref{R}) and (\ref{hij}) into (\ref{ahcond}), we obtain the equation of apparent horizons \cite{Nolan:1998,Kaloper:2010ec,Faraoni:2012gz}
\begin{equation}\label{HRZN}
R_{A}-2 m-H^2 {R_A^3} = 0 \,,
\end{equation}
where $H=H(t):=\dot a/a$ is the Hubble parameter and the dot stands for the time derivative. We only consider an expanding universe, i.e.\ $H>0$.
If $m H>1/3\sqrt{3}$, the above equation (\ref{HRZN}) has one real negative root and another two complex roots, which are all obviously unphysical. Thus in this paper we assume that $0<m H< 1/3\sqrt{3}$, then there are three real roots \cite{Faraoni:2012gz,Gregoris:2019oxz}
\begin{eqnarray}\label{eq:horizons}
R_{A1}=\frac{2}{\sqrt{3}H}\sin\vartheta \,,~~~~~~~~
R_{A2}=\frac{2}{\sqrt{3}H}\sin(\vartheta+\frac{2 \pi}{3}) \,,~~~~~~~~
R_{A3}=\frac{2}{\sqrt{3}H}\sin(\vartheta-\frac{2 \pi}{3}) \,,
\end{eqnarray}
where $\vartheta$ is defined by $\sin(3\vartheta)=3\sqrt{3}mH$, $\vartheta\in(0,\pi/3]$. One can prove that in this situation the first two roots above are always real and positive and the last one is negative (hence ignored), so there are two apparent horizons -- we denote the smaller one by $R_{Ab}$ and the larger one by $R_{Ac}$, which are the black hole and cosmological apparent horizons, respectively. Furthermore, one can prove that $2m<R_{A1}=R_{Ab}<3m$ and $R_{A2}=R_{Ac}>3m$ \cite{mass2,Lake:2011ni,Nolan:2014maa}. Note that we will ignore the extremal case that the two horizons degenerate at $R_{A1}=R_{A2}=3m$ with $mH=1/3\sqrt{3}$, since the temperature goes to zero in this situation, as can be seen later from (\ref{T1B}). Furthermore the derivative of (\ref{HRZN}) with respect to $t$ gives
\begin{equation}\label{dr}
\dot{R}_A=\frac{2 H \dot{H} R_{A}^{3}}{1-3 H^2 R_{A}^2}\,,
\end{equation}
which will be used later.

\bigskip

In this paper, we will only focus on the thermodynamics for the black hole, and we put the discussion for the cosmological horizon in Appendix \ref{A}. Hereinafter we drop the label ``$b$'' everywhere, i.e.\ from now on we take $R_A=R_{Ab}$.

\bigskip

Now we come to the thermodynamics. For dynamical spacetime, the surface gravity $\kappa$ is defined by \cite{Akbar:2006kj,Cai:2008ys,Cai:2008gw}
\begin{equation}\label{Kap}
\kappa=\frac{1}{2\sqrt{-h}}\frac{\partial}{\partial x^{i}}\left(\sqrt{-h}~h^{i j}\frac{\partial R}{\partial x^{j}}\right)\,,
\end{equation}
where $h=det(h_{i j})$.
Then, from (\ref{Kap}) by using (\ref{R}), (\ref{HRZN}) and (\ref{dr}), one can easily obtain the surface gravity on the horizon\footnote{Obviously, if $a(t)=$ const, it is just the surface gravity $\kappa=1/4m$ of the Schwarzschild black hole \cite{Gibbons:1977mu,Saida:2007ru}; if $\dot{R}_A=0$, it is reduces to the surface gravity of Schwarzschild-de Sitter black hole \cite{Urano:2009xn,Bhattacharya:2013tq,Kubiznak:2015bya}; and if $m=0$, it reduces to the surface gravity on the apparent horizon $R_{A}$ of the spatial flat FRW universe $\kappa=-[1-{\dot{R}_A}/{(2H R_A)}]/{R_A}$ \cite{Akbar:2006kj}.
}\cite{Ke-Xia:2009tzo,Akbar:2017ljn}:
\begin{equation}\label{Kappa}
\kappa|_{R=R_A}= \frac{1}{R_A}\left(\frac{3m}{R_A}-1\right) \left(1-\frac{{\dot{R}_A}}{2 H{}\sqrt{{R_{A}}(R_{A}-2 m)}}\right)\,.
\end{equation}
In order for the surface gravity to be positive at the black hole horizon, the following condition on the perfect fluid is required:
\begin{equation}
\frac{\rho}{3}-p_A< \frac{m}{2\pi R_{A}^3}\ ,\label{kcond}
\end{equation}
where $p_A=p|_{R=R_A}$ (see Appendix \ref{B} for the derivation).
Then the Hawking temperature is given by
\begin{equation}\label{T1B}
T=\frac{\kappa|_{R=R_A}}{2\pi}=\frac{1}{2 \pi R_{A}}\left(\frac{3m}{R_{A}}-1\right) \left(1-\frac{\dot{R}_{A}}{2 H{}\sqrt{R_{A}(R_{A}-2 m)}}\right)\,.
\end{equation}

Now we turn our attention to the first law of thermodynamics. In light of the work \cite{Hayward:1997jp,Cai:2005ra,Cai:2006rs,Gong:2007md,Hu:2015xva},
we write down the unified first law:
\begin{equation}\label{UFL}
d \tilde M=\tilde A\tilde\Psi+\tilde W d \tilde V \,,
\end{equation}
where $\tilde A=4\pi R^2$ is the surface area; $\tilde V=\frac43\pi R^3$ is the volume; $\tilde M$ is the Misner-Sharp energy of the region inside a radius $R$:
\begin{equation} \label{MSE}
\tilde{M}=\frac{R}{2}(1-h^{i j}\partial _{i}R\partial _{j}R)=m+\frac{1}{2}H^2 R^3\,,
\end{equation}
\begin{equation}
d \tilde M= H \dot H R^3 dt +\frac32 H^2 R^2 dR = -\tilde A({\rho}+p)H\sqrt{R(R-2 m)}d t+\tilde A{\rho} d R\,;
\end{equation}
$\tilde W$ is the work density
\begin{equation}\label{WPsi}
\tilde{W}=-\frac{1}{2}h^{i j}T_{i j}=\frac{1}{2}({\rho}-p) \,;
\end{equation}
$\tilde \Psi$ is the energy-supply:
\begin{equation}\label{WPsi1}
\tilde{\Psi}=(T_{i}^j{\partial _{j}R}+W{\partial _{i}R}) d{x^i}=\frac{1}{2}({\rho}+p)\left[-2H \sqrt{R(R-2 m)}d t+dR\right]\,.
\end{equation}

By restricting the unified first law on the black hole horizon, we obtain the first law of thermodynamics for the black hole:
\begin{eqnarray} \label{FAHB}
	d M = T d S + W d V \,,
\end{eqnarray}
where
\begin{eqnarray}
	M=\tilde M|_{R=R_A}=\frac{R_A}{2}\ , \ \ S=\frac14 \tilde A|_{R=R_A} = \pi R_{A}^2 \ , \ \
    V=\tilde V|_{R=R_A} =\frac43\pi R_A^3 \ , \ \ W=\tilde W|_{R=R_A}=\frac12(\rho-p_A) \,.\label{MSVWdef}
\end{eqnarray}

\bigskip

\section{Thermodynamic Equation of State of McVittie Black Hole} \label{sIII}

In order to investigate the Hawking-Page-like phase transition of McVittie black hole, we first derive its thermodynamic equation of state. A key point of constructing the equation of state is to find a proper definition of the thermodynamic pressure $P$. In this paper, it is natural to extract the definition from the first law (\ref{FAHB}), i.e.\ we define
\begin{equation}\label{BHWP}
	P:=-W=\frac{1}{2}(p_A-\rho)\,,
\end{equation}
so that, with $U:=M$, the first law can be rewritten as the usual form
\begin{equation}
dU=TdS-P dV\,. \label{newfl}
\end{equation}
From this equation, we can see that the volume defined in (\ref{MSVWdef}) is also the thermodynamic volume, i.e.\ conjugate to the thermodynamic pressure.

By using the following equations
\begin{equation}\label{rho1}
	\rho=\frac{3}{8 \pi}\left(\frac{1}{R_{A}^2}-\frac{2m}{R_{A}^3}\right) \,,
\end{equation}
\begin{eqnarray}\label{P}
	p_A=-\frac{3}{8 \pi}\left(\frac{1}{R_{A}^2}-\frac{2m}{R_{A}^3}\right)+\frac{(R_{A}-3m)\dot{R}_A}{4\pi H R_A^3 {\sqrt{R_A(R_A-2 m)}}}\,,
\end{eqnarray}
which are derived from the Friedmann's equations (\ref{rho1}) and (\ref{P}), and using (\ref{T1B}) to eliminate $\dot R_A$, we obtain the equation of state
\begin{eqnarray}\label{PVTB}
	P(T,R_A)=\frac{T}{2R_{A}}-\frac{1}{8\pi R_{A}^2} \,.
\end{eqnarray}

\begin{figure}[h]
 \begin{minipage}[t]{1\linewidth}
\includegraphics[width=7cm]{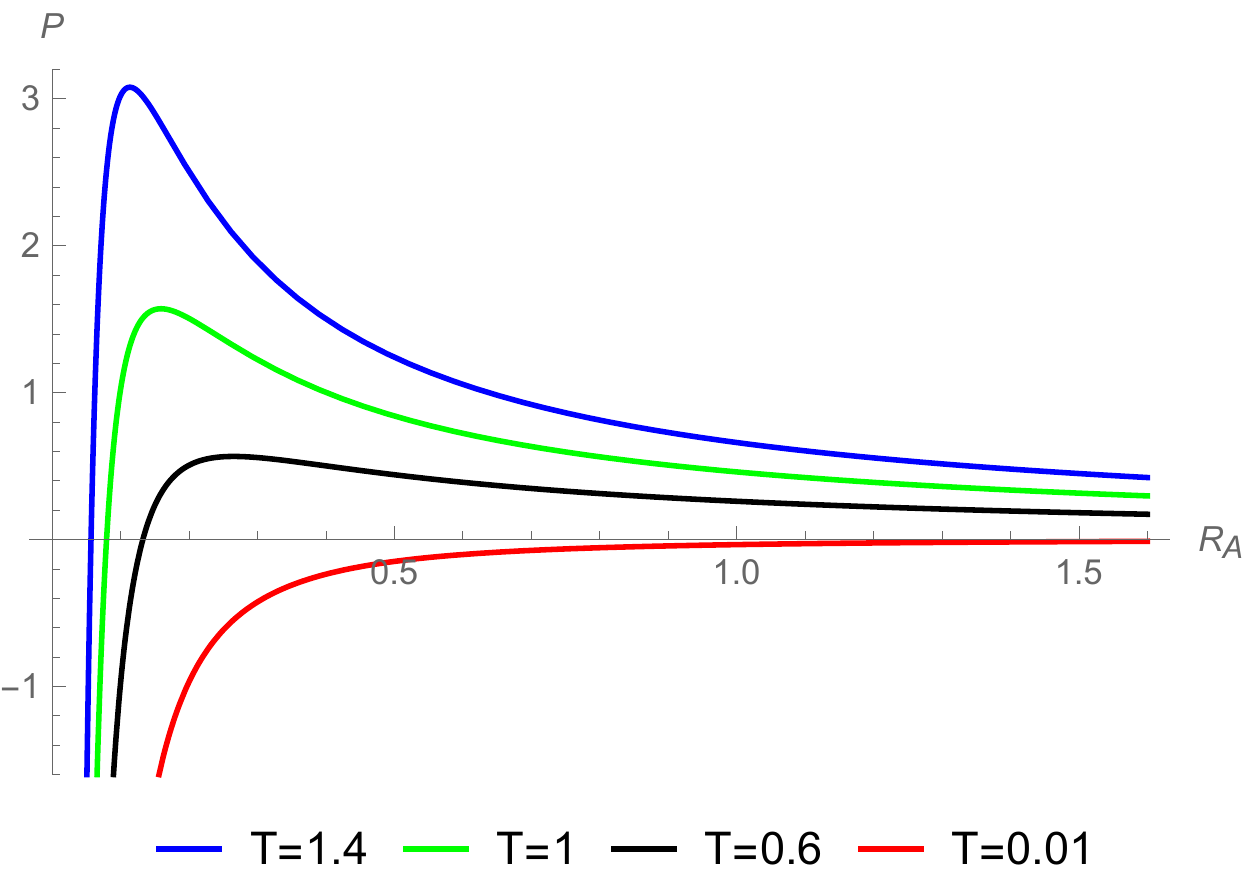}
\end{minipage}
\caption{\footnotesize The isothermal curves in the $P$-$R_A$ phase diagram.}
\label{BHPVGT}
\end{figure}

Before investigating the HP-like phase transition, let us first briefly demonstrate a few properties of this equation of state, whose isothermal curves are plotted in Fig.\ref{BHPVGT}.
\newline As can be seen from the figure, there is no van der Waals-like ($P$-$V$) phase transition, which can be rigorously proved in the following. The necessary condition for the existence of the $P$-$V$ phase transition is that \cite{Kubiznak:2012wp,Wei:2012ui,Cai:2013qga,Hu:2018qsy}
\begin{equation}\label{PV}
\left(\frac{\partial P}{\partial V}\right)_{T}=\left(\frac{\partial^2 P}{\partial V^2}\right)_{T}=0\,,
\end{equation}
or equivalently in this paper
\begin{equation}\label{PVTc}
\left(\frac{\partial P}{\partial R_A}\right)_{T}=\left(\frac{\partial^2 P}{\partial R^2_A}\right)_{T}=0\,,
\end{equation}
has a critical-point solution.
By substituting the equation of state (\ref{PVTB}) into (\ref{PVTc}), we obtain
\begin{equation}\label{PR12}
 \left(\frac{\partial P}{\partial R_A}\right)_{T}=\frac{1-2\pi R_A T}{4\pi R_A^3}=0\,,~~~~~~~~~~ \left(\frac{\partial^2 P}{\partial R^2_A}\right)_{T}=\frac{4\pi R_A T-3}{4\pi R_A^4}=0 \,.
\end{equation}
One can easily check that this kind of solution does not exist, and hence there is no $P$-$V$ phase transition\footnote{This conclusion is natural in the sense that the equation of state (\ref{PVTB}) resembles that of Schwarzschild-(A)dS black hole, which also has no $P$-$V$ phase transition \cite{Altamirano:2014tva,Ma:2015llh,Hansen:2016ayo}.}. Furthermore, from the first equation in (\ref{PR12}), one can see that when $R_A<\frac{1}{2\pi T}$, larger pressure leads to larger volume, i.e.\ $\left(\frac{\partial P}{\partial R_A}\right)_{T}>0$, and thus the corresponding black hole system is unstable.

\section{Hawking-Page-Like Phase Transition of the McVittie Black Hole}\label{sIV}

The Gibbs free energy is a crucial thermodynamic function for studying the HP(-like) phase transitions \cite{Hawking:1982dh,Spallucci:2013jja,Su:2019gby,Mbarek:2018bau}. For the McVittie black hole, the Gibbs free energy is given by
\begin{equation}\label{BHG}
 G = U+P V-TS=\frac{R_{A}}{12}(3-8\pi R_{A}^2 P) \,,
\end{equation}
where the thermodynamic equation of state (\ref{PVTB}) has been used. Note that a vanishing black hole has zero Gibbs free energy, as $G|_{R_A=0}=0$, which provides a reference background.

\begin{figure}[h]
\begin{minipage}[t]{7cm}
\centering
\includegraphics[width=7cm,height =6cm]{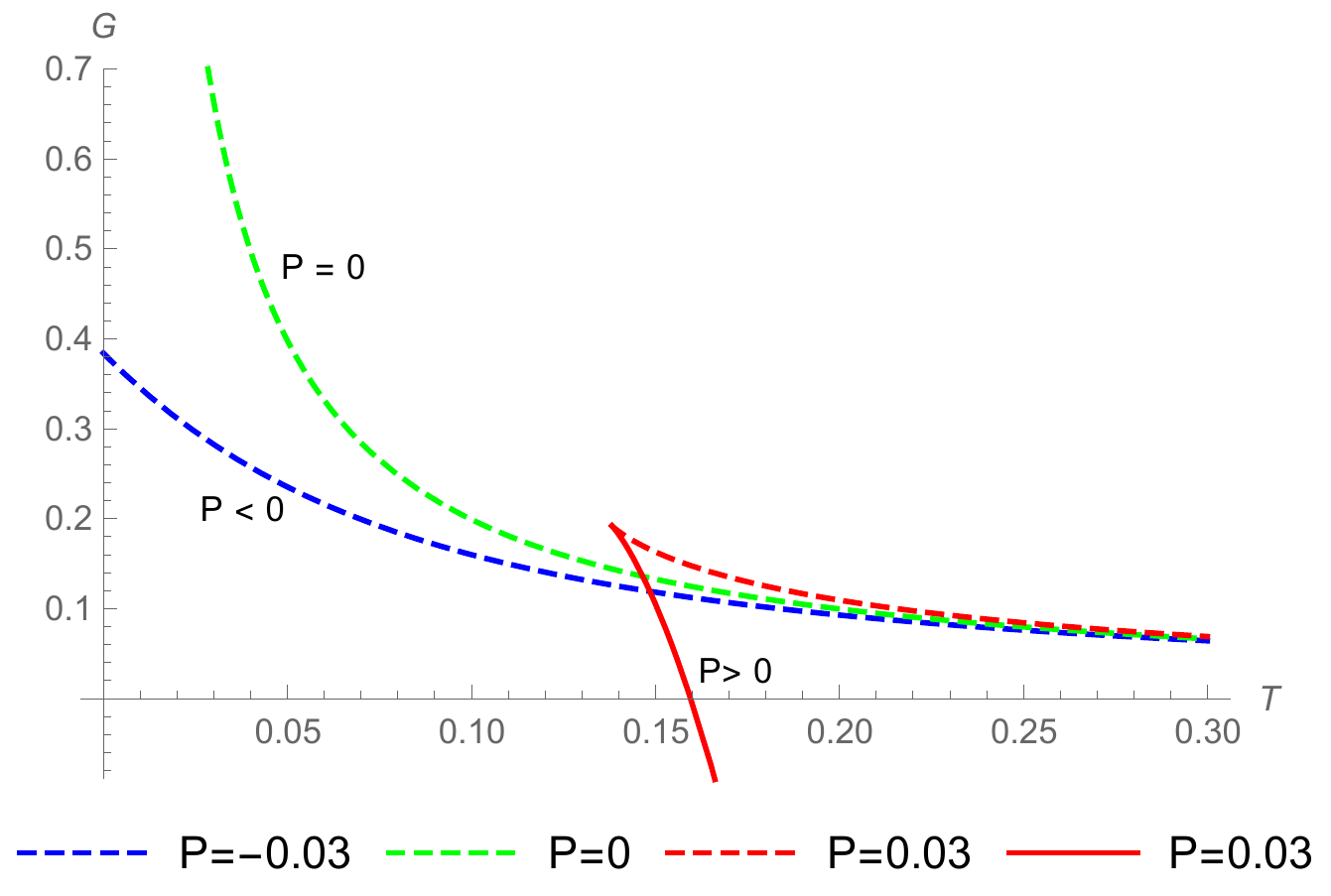}
\end{minipage}
\caption{The different behaviors of the isobaric curves for $P<0$, $P=0$ and $P>0$ in the $G$-$T$ phase diagram.
Only in the $P>0$ case, the isobaric curve crosses the horizontal $T$-axis, which means that there is a HP-like phase transition.}
{\label{THPGT1}}
\end{figure}

For the convenience of investigating the HP-like transition between the McVittie black hole and the FRW spacetime, we would like to rewrite the Gibbs free energy as a function $G=G(T,P)$. For this purpose we need to solve the equation of state (\ref{PVTB}) for $R_A$, and there are three different scenarios depending on the sign of $P$.
When $P=0$,
\begin{equation}
R_A=\frac{1}{4\pi T}
\end{equation}
is the only solution, and then
\begin{equation}
G=\frac{1}{16\pi T} \,.
\end{equation}
When $P>0$, there are two real positive solutions
\begin{equation}\label{RAminusP}
R_{A1^\prime}= \frac{1}{2\pi T+2\sqrt{\pi}\sqrt{-2P+\pi T^2}} \text{\ \ \ and\ \ \ } R_{A2^\prime}= \frac{1}{2\pi T-2\sqrt{\pi}\sqrt{-2P+\pi T^2}}\,,
\end{equation}
where obviously $R_{A1^\prime}<R_{A2^\prime}$, and their corresponding Gibbs free energies are respectively
\begin{equation}\label{SBHG}
 G(T,P)=\frac{(T-\sqrt{T^2-{2P}/{\pi}})[4P +\sqrt{\pi}T(\sqrt{\pi T^2-2P}-\sqrt{\pi}T)]}{48P^2}\,,
\end{equation}
and
\begin{equation}\label{LBHG}
 G(T,P)=\frac{(T+\sqrt{T^2-{2P}/{\pi}})[4P -\sqrt{\pi}T(\sqrt{\pi T^2-2P}+\sqrt{\pi}T)]}{48P^2}\,.
\end{equation}
For $P<0$, the solutions are formally the same as (\ref{RAminusP}), but only $R_{A1^\prime}$ is positive, while $R_{A2^\prime}$ is negative and hence unphysical, so we should only use the former expression of $G(T,P)$. We plot the isobaric curves of Gibbs free energy in Fig.\ref{THPGT1}. As indicated in the figure, the HP-like phase transition happens only in the scenario $P>0$, which we will elaborate below.

\begin{figure}[h]
\begin{minipage}[t]{7cm}
\centering
\includegraphics[width=8cm,height =5cm]{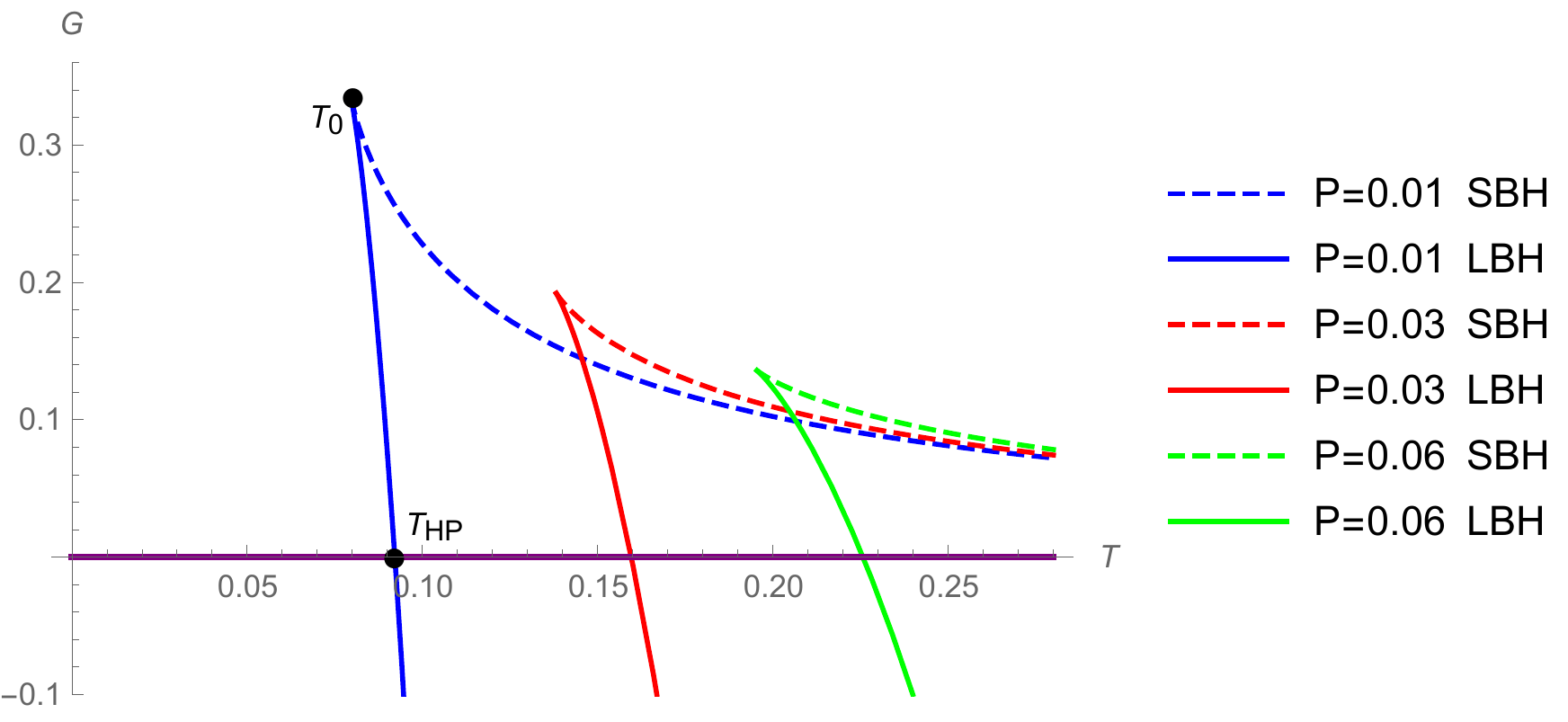}
\end{minipage}
\caption{The isobaric curves in the $G$-$T$ phase diagram. The curves for the small and large black holes meet at the minimum temperature $T_0$.
The Gibbs free energy of the small black hole is always above that of the large black hole and the $T$-axis, so it is always unstable compared with the large black hole and the FRW background. The isobaric curves of the large black hole and the FRW background have an intersection at the HP temperature $T_{HP}$, which corresponds to a first-order phase transition.}
{\label{THPGT}}
\end{figure}

In the $P>0$ case, we plot in Fig. \ref{THPGT} a few isobaric curves of the Gibbs free energy of the black hole in comparison a vanishing black hole in the FRW spacetime. As illustrated in the figure, the smaller black hole always has higher Gibbs free energy than the FRW spacetime, and hence the former is relatively unstable. Regarding the larger black hole, the temperature of HP-like phase transition is obtained by solving $G(T,P)=0$ for $T$, which gives
\begin{equation}\label{THP}
 T_{\text{HP}}=\sqrt{\frac{8 P}{3\pi}}\,.
\end{equation}
Obviously, when $T<T_{\text{HP}}$ the black hole has positive Gibbs free energy, then the FRW spacetime is more stable. On the contrary, when $T>T_{\text{HP}}$ the black hole has negative Gibbs free, and thus it is more stable, which means a HP-like phase transition from FRW spacetime to a McVittie black hole.

\begin{figure}[h]
\centering
\begin{minipage}[t]{7cm}
\centering
\includegraphics[width=6.5cm,height =5cm]{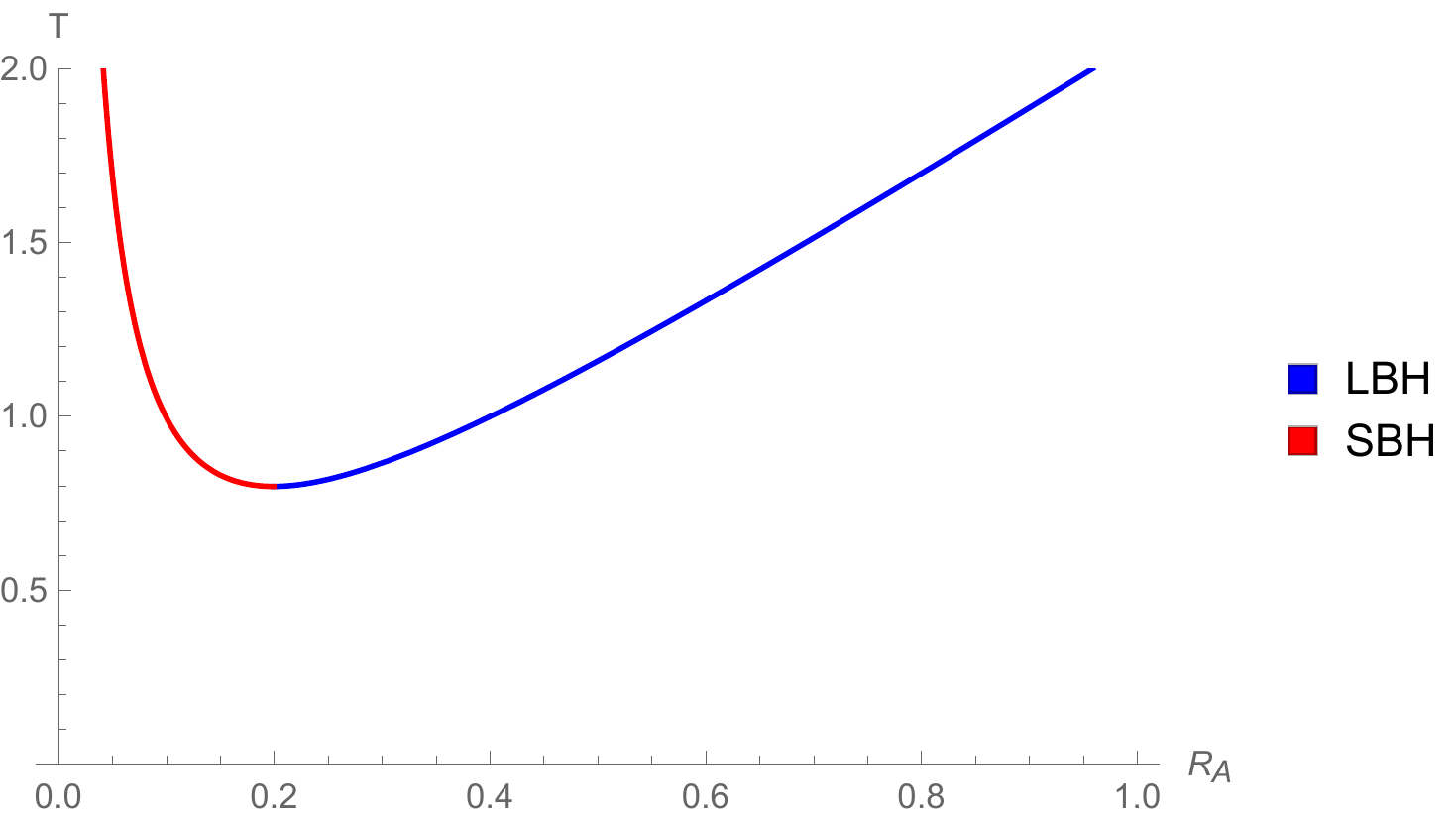}
\end{minipage}
\caption{An isobaric curve ($P=1$) in the $T$-$R_{A}$ phase diagram of the black hole. The temperature and horizon radius figure indicates the existence of a minimum temperature $T_0$. For $T < T_0$, no black holes exist, except the FRW background. For $T > T_0$, there are two phases: the small black hole (shown by red curve) and the large one (shown by blue curve).}{\label{BHTsR}}
\end{figure}

Note that in Fig.\ref{THPGT}, the larger and the smaller black holes coincide at the temperature $T_0$, which is better presented in Fig.\ref{BHTsR}, as the minimum point on an isobaric $T$-$R_A$ line plotted according to
\begin{equation}\label{TVPB}
  T=2P R_{A}+\frac{1}{4\pi R_{A}}
\end{equation}
derived from the equation of state (\ref{PVTB})\,.
It can be easily calculated that
\begin{equation}\label{T0}
 T_0= \sqrt{\frac{2 P}{\pi}}\,,
\end{equation}
below which no black hole exists.
Interestingly, the ratio between $T_{\text{HP}}$ and $T_0$ is a constant
\begin{equation}\label{THPT0}
 \frac{T_{\text{HP}}}{T_0}=\frac{2}{3} \sqrt{3}\,,
\end{equation}
which coincides with that of Schwarzschild-AdS black hole \cite{Kubiznak:2016qmn,Wei:2020kra}.

\section{Conclusions and Discussions}\label{sVI}

In this paper, we have investigated the thermodynamic properties of the McVittie spacetime in the framework of Einstein's general relativity. We have focused on the thermodynamics associated with the black hole horizon (while that for the cosmological horizon is in Appendix \ref{A}). We have formulated its first law of thermodynamics, using the Misner-Sharp energy, the Hawking temperature, the Bekenstein-Hawking entropy, the thermodynamic volume and the thermodynamic pressure defined as $P:=-W$, where $W$ is the work density of the perfect fluid. We have also formulated the equation of state that connects the thermodynamic pressure, the Hawking temperature and the horizon radius. With this setup, we have investigated the Gibbs free energy for the black hole in comparison with that for the FRW spacetime. In the case $P>0$, we have found the Hawking-Page-like phase transition, which means the transition between the McVittie black hole and thermal radiation in FRW spacetime. As far as we know, this is the first time that the Hawking-Page-like phase transition is discovered on FRW background.

In this paper, the thermodynamic pressure is $P=(p_A-\rho)/2$, where $p_A$ and $\rho$ stand for the pressure and energy density of the perfect fluid at the horizon. It is interesting to notice that the Hawking-Page-like phase transition exists only when $P>0$, i.e.\  $p_A>\rho$, which violates the dominant energy condition \cite{Kontou:2020bta}. Since classical matter in theory never violates this condition, our result could be an implication that quantum effects of matter field or modification to general relativity might be crucial for this transition to happen, which is worth studying in the future.
An interesting scenario where our result might be applicable is the early universe that is both hot and fast expanding. In this scenario, the environment might be hotter than the Hawking-Page temperature and thus the black hole phase is stable. Later, as the universe cools down below the Hawking-Page temperature, black holes become unstable and can evaporate into thermal gas. This process might affect the evolution of our universe. The work in this paper has laid a foundation to investigate the detailed mechanism of how primordial black holes were created and evolved from the thermodynamic point of view, which we will leave for our future research.

\appendix
\section{First Law of Thermodynamics and Equation of State for the Cosmological Apparent Horizon}\label{A}

In this appendix, we study the first law of thermodynamics for the cosmological apparent horizon of the McVittie spacetime and further derive its thermodynamic equation of state.

The surface gravity of the cosmological apparent horizon is negative, see corresponding analyses in Appendix \ref{B}. Using Eq.(\ref{Kappa}), the Hawking temperature of the cosmological apparent horizon reads
\begin{equation}\label{T1C}
T|_{R=R_{Ac}}=-\frac{\kappa |_{R=R_{Ac}}}{2\pi}=\frac{1}{2 \pi R_{Ac}}\left(1 -\frac{3m}{R_{Ac}}\right) \left(1-\frac{\dot{R}_{Ac}}{2 H{}\sqrt{R_{Ac}(R_{Ac}-2 m)}}\right)\,.
\end{equation} 	
From (\ref{WPsi}), one obtains the work density at the cosmological apparent horizon
\begin{eqnarray}\label{Wl}
W_c=\frac{T_c}{2R_{Ac}}+\frac{1}{8\pi R_{Ac}^2}\,.
\end{eqnarray}
From the unified first law (\ref{UFL}), one can also derive the first law of thermodynamics for the cosmological horizon
\begin{eqnarray} \label{FAHC}
d M_c=\frac{\kappa_c}{8\pi}d A_{c}+W_{c} d V_{c}=-T_{c}d S_{c}+W_{c} d V_{c} \,,
\end{eqnarray}
where $S_c=A_c/4=\pi R_{Ac}^2$, $V_c=4\pi R_{Ac}^3/3$ are the Bekenstein-Hawking entropy and thermodynamic volume, respectively. Note that there is a minus sign in front of the first term on the r.h.s of (\ref{FAHC}), caused by the negative surface gravity $\kappa_c$ \cite{Akbar:2006kj,Dolan:2013ft,Abdusattar:2021wfv}, which is different from the standard first law of thermodynamics $dU=TdS-P dV$.
Therefore,
one should identify the internal energy $U$ as the opposite of the Misner-Sharp energy $M_c$, i.e. $U:=-M_c$ in (\ref{FAHC}), and the thermodynamic pressure $P$ should be identified with the work density $W_c$
\begin{equation}\label{CWP}
P:=W_c\,.
\end{equation}
The equation of state can be obtained directly from (\ref{Wl}) and (\ref{CWP})
\begin{eqnarray}\label{PVTC}
P=\frac{T_c}{2R_{Ac}}+\frac{1}{8\pi R_{Ac}^2}\,.
\end{eqnarray}
It is interesting to note that, Eq.(\ref{PVTC}) is formally the same as that for the FRW universe in Einstein gravity discussed in \cite{Kong:2021qiu}. This implies that (\ref{PVTC}) might be a model-independent expression.
Using the same methods as those in the main text, here we do not find van der Waals-like or Hawking-Page-like phase transitions.

\bigskip

\section{The Positivity Condition of Surface Gravity} \label{B}

In this appendix, we prove that the expression (\ref{Kappa}) being positive i.e.\ $\kappa|_{R=R_A}>0$ is equivalent to (\ref{kcond}).

Because we have $2m<R_{A}<3m$, it is obvious that  $\kappa|_{R=R_A}>0$  is equivalent to
\begin{equation}
\frac{\dot{R}_{A}}{{2 H{}\sqrt{R_{A}(R_{A}-2 m)}}}<1 \,,
\end{equation}
which, by substituting (\ref{HRZN}) and (\ref{dr}), can be converted into
\begin{equation}
\frac{R_A^4 \dot{H}}{2(3m-R_A){}\sqrt{R_{A}(R_{A}-2 m)}}<1 \,,
\end{equation}
and by substituting (\ref{GRTp}) is further equivalent to
\begin{eqnarray} \label{kappaA1}
\frac{2\pi (\rho+p_A)R_A^3}{R_A-3m}<1\,,
\end{eqnarray}
and hence
\begin{eqnarray} \label{kappaA2}
\rho+p_A>\frac{R_A-3m}{2\pi R_A^3} \,.
\end{eqnarray}
Moreover, rewriting the above formula as
\begin{eqnarray} \label{kappaA3}
\rho+p_A>\frac{1}{2\pi}\left(\frac{1}{R_A^2}-\frac{2m}{R_A^3}\right)-\frac{m}{2\pi R_A^3} \,.
\end{eqnarray}
and using (\ref{rho1}) one can prove that it is equivalent to
\begin{eqnarray} \label{kappaA4}
\rho+p_A>\frac{1}{2\pi}\frac{8\pi}{3}\rho-\frac{m}{2\pi R_A^3} \,.
\end{eqnarray}
i.e.\
\begin{eqnarray} \label{kappaAb}
\frac{\rho}{3}-p_A<\frac{m}{2\pi R_A^3} \,,
\end{eqnarray}
as in (\ref{kcond}). The physical implication of (\ref{kappaAb}) is not yet fully understood, but its $m\rightarrow0$ limit is clear. Because $2m<R_A<3m$, in the $m\rightarrow0$ limit, the r.h.s.\ goes to infinity and hence such a condition is always satisfied.

\bigskip
As an additional remark, a similar condition can be derived for the surface gravity being negative on the cosmological horizon (we denote its radius by $R_{Ac}$ here).
In this situation, (\ref{kappaA1}) still holds by replacing $R_{A}$ with $R_{Ac}$. Then by taking note that $R_{Ac}>3m$, one can prove that
\begin{eqnarray} \label{kappaA4}
\frac{\rho}{3}-p_c> \frac{m}{2\pi R_{Ac}^3}\,.
\end{eqnarray}
is equivalent to $\kappa|_{R=R_{Ac}}<0$, where $p_c=p|_{R=R_{Ac}}$. In the $m\rightarrow0$ limit, $R_{Ac}\rightarrow\frac{1}{H}$, and hence (\ref{kappaA4}) becomes $\frac{\rho}{3}-p_c>0$, which is consistent with that in FRW universe.

\subsection*{Acknowledgement}

We are grateful to Profs. Hongsheng Zhang, Yen Chin Ong, Xiao-Mei Kuang, Li-Ming Cao and Drs. Arshad Ali, Tao-Tao Sui for interesting and stimulating discussions. This work is supported by National Natural Science Foundation of China (NSFC) under grants Nos. 12175105, 11575083, 11565017. Top-notch Academic Programs Project of Jiangsu Higher Education Institutions (TAPP), and is also supported by \textquotedblleft the Fundamental Research Funds for the Central Universities, NO. NS2020054\textquotedblright\ of China.

{}


\begin{thebibliography}{99}


%

\bibitem{Sasaki:2016jop}
M.~Sasaki, T.~Suyama, T.~Tanaka and S.~Yokoyama,
``Primordial Black Hole Scenario for the Gravitational-Wave Event GW150914'',
{\hypersetup{urlcolor=vividviolet}\href{https://inspirehep.net/literature/1435028}{Phys. Rev. Lett. \textbf{117} (2016) no.6, 061101} {[erratum: Phys. Rev. Lett. \textbf{121} (2018) no.5, 059901]}}, \href{https://arxiv.org/abs/1603.08338}{[arXiv:astro-ph.CO/1603.08338]}


\bibitem{Niemeyer:1999ak}
J.~C.~Niemeyer and K.~Jedamzik,
``Dynamics of primordial black hole formation'',
{\hypersetup{urlcolor=vividviolet}\href{https://inspirehep.net/literature/494294}{Phys. Rev. D \textbf{59} (1999), 124013}},\href{https://arxiv.org/abs/astro-ph/9901292}{[arXiv:astro-ph/9901292]};
T.~Harada and B.~J.~Carr,
{\hypersetup{urlcolor=vividviolet}\href{https://inspirehep.net/literature/666223}{Phys. Rev. D \textbf{71} (2005), 104009}}, \href{https://arxiv.org/abs/astro-ph/0412134}{[arXiv:astro-ph/0412134]};
B.~Carr, F.~Kuhnel and M.~Sandstad,
{\hypersetup{urlcolor=vividviolet}\href{https://inspirehep.net/literature/1477198}{Phys. Rev. D \textbf{94} (2016) no.8, 083504}},\href{https://arxiv.org/abs/1607.06077}{[arXiv:astro-ph.CO/1607.06077]}.


\bibitem{Carr:1974nx}
B.~J.~Carr and S.~W.~Hawking,
``Black holes in the early Universe'',
{\hypersetup{urlcolor=vividviolet}\href{https://inspirehep.net/literature/95453}{Mon. Not. Roy. Astron. Soc. \textbf{168} (1974), 399-415}}.

\bibitem{McVittie:1933zz}
G.~C.~McVittie, ``The Mass-Particle in an Expanding Universe",
{\hypersetup{urlcolor=vividviolet}\href{https://academic.oup.com/mnras/article/93/5/325/951988}{Mon. Not. Roy. Astron. Soc. \textbf{93}, 325-339 (1933)}}.

\bibitem{Nolan:1998}
Brien C. Nolan, ``A point mass in an isotropic universe: Existence, uniqueness, and basic properties", {\hypersetup{urlcolor=vividviolet}\href{https://journals.aps.org/prd/abstract/10.1103/PhysRevD.58.064006}{Phys. Rev. D \textbf{58} (1998) 064006}}, \href{https://arxiv.org/abs/gr-qc/9805041}{[arXiv:gr-qc/9805041]}.

\bibitem{Kaloper:2010ec}
Nemanja Kaloper, Matthew Kleban, Damien Martin, ``McVittie's Legacy: Black Holes in an Expanding Universe", {\hypersetup{urlcolor=vividviolet}\href{https://journals.aps.org/prd/abstract/10.1103/PhysRevD.81.104044}{Phys. Rev. D \textbf{81} (2010) 104044}}, \href{https://arxiv.org/abs/1003.4777}{[arXiv:hep-th/1003.4777]}.

\bibitem{daSilva:2012nh}
A.~M.~da Silva, M.~Fontanini and D.~C.~Guariento,
``How the expansion of the universe determines the causal structure of McVittie spacetimes'',
{\hypersetup{urlcolor=vividviolet}\href{https://inspirehep.net/literature/1205068}{Phys. Rev. D \textbf{87} (2013) no.6, 064030}}, \href{https://arxiv.org/abs/1212.0155}{[arXiv:gr-qc/1212.0155]}.

\bibitem{Faraoni:2012gz}
V.~Faraoni, A.~F.~Zambrano Moreno and R.~Nandra,
``Making sense of the bizarre behaviour of horizons in the McVittie spacetime",
{\hypersetup{urlcolor=vividviolet}\href{https://inspirehep.net/literature/1087339}{Phys. Rev. D \textbf{85}, 083526 (2012)}}, \href{https://arxiv.org/pdf/1202.0719}{[arXiv:gr-qc/1202.0719]}.

\bibitem{Faraoni:2015ula}
V.~Faraoni,
``Cosmological and Black Hole Apparent Horizons'',
{\hypersetup{urlcolor=vividviolet}\href{https://inspirehep.net/literature/1382050}{Lect. Notes Phys. \textbf{907}, pp.1-199 (2015)}}.

\bibitem{Antoniou:2016obw}
I.~Antoniou and L.~Perivolaropoulos,
``Geodesics of McVittie Spacetime with a Phantom Cosmological Background'',
{\hypersetup{urlcolor=vividviolet}\href{https://inspirehep.net/literature/1426671}{Phys. Rev. D \textbf{93}, no.12, 123520 (2016)}}, \href{https://arxiv.org/abs/1603.02569}{[arXiv:gr-qc/1603.02569]}.

\bibitem{Piattella:2015xga}
O.~F.~Piattella,
``Lensing in the McVittie metric'',
{\hypersetup{urlcolor=vividviolet}\href{https://inspirehep.net/literature/1388683}{Phys. Rev. D \textbf{93}, no.2, 024020 (2016)}{[erratum: Phys. Rev. D \textbf{93}, no.12, 129901 (2016)]}}, \href{https://arxiv.org/abs/1508.04763}{[arXiv:astro-ph.CO/1508.04763]}.

\bibitem{Gregoris:2019oxz}
D.~Gregoris, Y.~C.~Ong and B.~Wang,
``The Horizon of the McVittie Black Hole: On the Role of the Cosmic Fluid Modeling",
{\hypersetup{urlcolor=vividviolet}\href{https://inspirehep.net/literature/1763153}{Eur. Phys. J. C \textbf{80}, no.2, 159 (2020)}}, \href{https://arxiv.org/abs/1911.01809}{[arXiv:gr-qc/1911.01809]}.

\bibitem{Ruiz:2020yye}
F.~Ruiz, C.~Molina and J.~A.~S.~Lima,
``Dynamical model for primordial black holes'',
{\hypersetup{urlcolor=vividviolet}\href{https://inspirehep.net/literature/1830435}{Phys. Rev. D \textbf{102} (2020) no.12, 123516}}, \href{https://arxiv.org/abs/2011.07079}{[arXiv:gr-qc/2011.07079]}.


\bibitem{Hayward:1997jp}
S.~A.~Hayward,``Unified first law of black hole dynamics and relativistic thermodynamics",
{\hypersetup{urlcolor=vividviolet}\href{https://inspirehep.net/literature/449938}{Class. Quant. Grav. \textbf{15}, 3147-3162 (1998)}}, \href{https://arxiv.org/abs/gr-qc/9710089}{[arXiv:gr-qc/9710089]}.

\bibitem{Ke-Xia:2009tzo}
J.~Ke-Xia, K.~San-Min and P.~Dan-Tao,
``Hawking Radiation as tunneling and the unified first law of thermodynamics for a class of dynamical black holes", {\hypersetup{urlcolor=vividviolet}\href{https://inspirehep.net/literature/885830}{Int. J. Mod. Phys. D \textbf{18}, 1707-1717 (2009)}}, \href{https://arxiv.org/abs/1105.0595}{[arXiv:hep-th/1105.0595]}.

\bibitem{Akbar:2017ljn}
M.~Akbar, T.~Brahimi and S.~M.~Qaisar,
``Thermodynamic Analysis of Cosmological Black Hole",
{\hypersetup{urlcolor=vividviolet}\href{https://inspirehep.net/literature/1511765}{Commun. Theor. Phys. \textbf{67}, no.1, 47 (2017)}}.

\bibitem{DiCriscienzo:2007pcr}
R.~Di Criscienzo, M.~Nadalini, L.~Vanzo, S.~Zerbini and G.~Zoccatelli,
``On the Hawking radiation as tunneling for a class of dynamical black holes'',
{\hypersetup{urlcolor=vividviolet}\href{https://inspirehep.net/literature/756933}{Phys. Lett. B \textbf{657}, 107-111 (2007)}}, \href{https://arxiv.org/abs/0707.4425}{[arXiv:hep-th/0707.4425]};
R.~Di Criscienzo and L.~Vanzo,
{\hypersetup{urlcolor=vividviolet}\href{https://inspirehep.net/literature/780626}{EPL \textbf{82} (2008) no.6, 60001}}, \href{https://arxiv.org/abs/0803.0435}{[arXiv:hep-th/0803.0435]}.

\bibitem{Faraoni:2007gq}
V.~Faraoni,
``The Hawking temperature of expanding cosmological black holes'',
{\hypersetup{urlcolor=vividviolet}\href{https://inspirehep.net/literature/763854}{Phys. Rev. D \textbf{76}, 104042 (2007)}}, \href{https://arxiv.org/abs/0710.2122}{[arXiv:gr-qc/0710.2122]}.


\bibitem{Kastor:2009wy}
D.~Kastor, S.~Ray and J.~Traschen,
``Enthalpy and the Mechanics of AdS Black Holes",
{\hypersetup{urlcolor=vividviolet}\href{https://inspirehep.net/literature/818236}{Class. Quant. Grav. \textbf{26}, 195011 (2009)}},\href{https://arxiv.org/pdf/0904.2765}{[arXiv:hep-th/0904.2765]}.

\bibitem{Dolan:2010ha}
B.~P.~Dolan,``The cosmological constant and the black hole equation of state",
{\hypersetup{urlcolor=vividviolet}\href{https://inspirehep.net/literature/866593}{Class. Quant. Grav. \textbf{28}, 125020 (2011)}}, \href{https://arxiv.org/pdf/1008.5023}{[arXiv:gr-qc/1008.5023]};
B.~P.~Dolan,
{\hypersetup{urlcolor=vividviolet}\href{https://inspirehep.net/literature/916548}{Class. Quant. Grav. \textbf{28}, 235017 (2011)}},\href{https://arxiv.org/abs/1106.6260}{[arXiv:gr-qc/1106.6260]}.

\bibitem{Cvetic:2010jb}
M.~Cvetic, G.~W.~Gibbons, D.~Kubiznak and C.~N.~Pope,
``Black Hole Enthalpy and an Entropy Inequality for the Thermodynamic Volume'',
{\hypersetup{urlcolor=vividviolet}\href{https://inspirehep.net/literature/881245}{Phys. Rev. D \textbf{84}, 024037 (2011)}},\href{https://arxiv.org/abs/1012.2888}{[arXiv:hep-th/1012.2888]}.

\bibitem{Kubiznak:2012wp}
D.~Kubiznak and R.~B.~Mann,
``$P$-$V$ criticality of charged AdS black holes",
{\hypersetup{urlcolor=vividviolet}\href{https://inspirehep.net/literature/1113435}{JHEP \textbf{07}, 033 (2012)}}, \href{https://arxiv.org/pdf/1205.0559}{[arXiv:hep-th/1205.0559]}.


\bibitem{Altamirano:2014tva}
N.~Altamirano, D.~Kubiznak, R.~B.~Mann and Z.~Sherkatghanad,
``Thermodynamics of rotating black holes and black rings: phase transitions and thermodynamic volume",
{\hypersetup{urlcolor=vividviolet}\href{https://inspirehep.net/literature/1276853}{Galaxies \textbf{2} (2014), 89-159}}, \href{https://arxiv.org/abs/1401.2586}{[arXiv:hep-th/1401.2586]};
D.~Kubiznak and R.~B.~Mann,
``Black hole chemistry",
{\hypersetup{urlcolor=vividviolet}\href{https://inspirehep.net/literature/1289053}{Can. J. Phys. \textbf{93}, no.9, 999-1002 (2015)}},\href{https://arxiv.org/abs/1404.2126}{[arXiv:gr-qc/1404.2126]}.


\bibitem{Kubiznak:2016qmn}
D.~Kubiznak, R.~B.~Mann and M.~Teo,
``Black hole chemistry: thermodynamics with Lambda'',
{\hypersetup{urlcolor=vividviolet}\href{https://inspirehep.net/literature/1482749}{Class. Quant. Grav. \textbf{34} (2017) no.6, 063001}}, \href{https://arxiv.org/abs/1608.06147}{[arXiv:hep-th/1608.06147]}.


\bibitem{Gunasekaran:2012dq}
S.~Gunasekaran, R.~B.~Mann and D.~Kubiznak,
``Extended phase space thermodynamics for charged and rotating black holes and Born-Infeld vacuum polarization",
{\hypersetup{urlcolor=vividviolet}\href{https://inspirehep.net/literature/1183854}{JHEP \textbf{11}, 110 (2012)}}, \href{https://arxiv.org/abs/1208.6251}{[arXiv:hep-th/1208.6251]}.

\bibitem{Wei:2012ui}
S.~W.~Wei and Y.~X.~Liu,
``Critical phenomena and thermodynamic geometry of charged Gauss-Bonnet AdS black holes",
{\hypersetup{urlcolor=vividviolet}\href{https://inspirehep.net/literature/1184929}{Phys. Rev. D \textbf{87}, no.4, 044014 (2013)
}}, \href{https://arxiv.org/abs/1209.1707}{[arXiv:gr-qc/1209.1707]};
S.~W.~Wei and Y.~X.~Liu,
{\hypersetup{urlcolor=vividviolet}\href{https://inspirehep.net/literature/1789085}{Phys. Rev. D \textbf{101}, no.10, 104018 (2020)}}, \href{https://arxiv.org/abs/2003.14275}{[arXiv:gr-qc/2003.14275]}.


\bibitem{Hendi:2012um}
S.~H.~Hendi and M.~H.~Vahidinia,
``Extended phase space thermodynamics and $P$-$V$ criticality of black holes with a nonlinear source'',
{\hypersetup{urlcolor=vividviolet}\href{https://inspirehep.net/literature/1208685}{Phys. Rev. D \textbf{88}, no.8, 084045 (2013)}},\href{https://arxiv.org/abs/1212.6128}{[arXiv:hep-th/1212.6128]};
S.~H.~Hendi, R.~B.~Mann, S.~Panahiyan and B.~Eslam Panah,
{\hypersetup{urlcolor=vividviolet}\href{https://inspirehep.net/literature/1711378}{Phys. Rev. D \textbf{95}, no.2, 021501 (2017)}},\href{https://arxiv.org/pdf/1812.09938}{[arXiv:gr-qc/1702.00432]}.

\bibitem{Altamirano:2013ane}
N.~Altamirano, D.~Kubiznak and R.~B.~Mann,
``Reentrant phase transitions in rotating anti\textendash{}de Sitter black holes'',
{\hypersetup{urlcolor=vividviolet}\href{https://inspirehep.net/literature/1239816}{Phys. Rev. D \textbf{88}, no.10, 101502 (2013)
}}, \href{https://arxiv.org/abs/1306.5756}{[arXiv:hep-th/1306.5756]}.


\bibitem{Cai:2013qga}
R.~G.~Cai, L.~M.~Cao, L.~Li and R.~Q.~Yang,
``$P$-$V$ criticality in the extended phase space of Gauss-Bonnet black holes in AdS space",
{\hypersetup{urlcolor=vividviolet}\href{https://inspirehep.net/literature/1239956}{JHEP \textbf{09}, 005 (2013)}}, \href{https://arxiv.org/abs/1306.6233}{[arXiv:gr-qc/1306.6233]};
R.~G.~Cai, Y.~P.~Hu, Q.~Y.~Pan and Y.~L.~Zhang,
{\hypersetup{urlcolor=vividviolet}\href{https://inspirehep.net/literature/1315428}{Phys. Rev. D \textbf{91}, no.2, 024032 (2015)
}}, \href{https://arxiv.org/abs/1409.2369}{[arXiv:hep-th/1409.2369]}.

\bibitem{Bhattacharya:2017nru}
K.~Bhattacharya, B.~R.~Majhi and S.~Samanta,
``Van der Waals criticality in AdS black holes: a phenomenological study'',
{\hypersetup{urlcolor=vividviolet}\href{https://inspirehep.net/literature/1622513}{Phys. Rev. D \textbf{96}, no.8, 084037 (2017)}}, \href{https://arxiv.org/abs/1709.02650}{[arXiv:gr-qc/1709.02650]}.

\bibitem{Hu:2018qsy}
Y.~P.~Hu, H.~A.~Zeng, Z.~M.~Jiang and H.~Zhang,
``$P$-$V$ criticality in the extended phase space of black holes in Einstein-Horndeski gravity",
{\hypersetup{urlcolor=vividviolet}\href{https://inspirehep.net/literature/1711378}{Phys. Rev. D \textbf{100}, no.8, 084004 (2019)}},\href{https://arxiv.org/pdf/1812.09938}{[arXiv:gr-qc/1812.09938]};
J.~Xu, L.~M.~Cao and Y.~P.~Hu,
{\hypersetup{urlcolor=vividviolet}\href{https://inspirehep.net/literature/1375807}{Phys. Rev. D \textbf{91}, no.12, 124033 (2015)}}, \href{https://arxiv.org/pdf/1506.03578}{[arXiv:gr-qc/1506.03578]};
Y.~P.~Hu, L.~Cai, X.~Liang, S.~B.~Kong and H.~Zhang,
{\hypersetup{urlcolor=vividviolet}\href{https://inspirehep.net/literature/1823798}{Phys. Lett. B \textbf{822}, 136661 (2021)}}, \href{https://arxiv.org/abs/2010.09363}{[arXiv:gr-qc/2010.09363]}.

\bibitem{Li:2020xkh}
R.~Li and J.~Wang,
``Hawking radiation and $P$-$V$ criticality of charged dynamical (Vaidya) black hole in anti-de Sitter space", {\hypersetup{urlcolor=vividviolet}\href{https://inspirehep.net/literature/1818176}{Phys. Lett. B \textbf{813}, 136035 (2021)}}, \href{https://arxiv.org/abs/2009.09319}{[arXiv:gr-qc/2009.09319]}.

\bibitem{Hawking:1982dh}
S.~W.~Hawking and D.~N.~Page,
``Thermodynamics of Black Holes in anti-de Sitter Space",
{\hypersetup{urlcolor=vividviolet}\href{https://inspirehep.net/literature/181925}{Commun. Math. Phys. \textbf{87}, 577 (1983)}}.

\bibitem{Cai:2007wz}
R.~G.~Cai, S.~P.~Kim and B.~Wang,
``Ricci flat black holes and Hawking-Page phase transition in Gauss-Bonnet gravity and dilaton gravity'',
{\hypersetup{urlcolor=vividviolet}\href{https://inspirehep.net/literature/750802}{Phys. Rev. D \textbf{76} (2007), 024011}}, \href{https://arxiv.org/abs/0705.2469}{[arXiv:hep-th/0705.2469]}.


\bibitem{Czinner:2017tjq}
V.~G.~Czinner and H.~Iguchi,
``Thermodynamics, stability and Hawking\textendash{}Page transition of Kerr black holes from R\'enyi statistics'',
{\hypersetup{urlcolor=vividviolet}\href{https://inspirehep.net/literature/1514023}{Eur. Phys. J. C \textbf{77}, no.12, 892 (2017)}}, \href{https://arxiv.org/abs/1702.05341}{[arXiv:gr-qc/1702.05341]}.

\bibitem{Mejrhit:2019oyi}
K.~Mejrhit and S.~E.~Ennadifi,
``Thermodynamics, stability and Hawking\textendash{}Page transition of black holes from non-extensive statistical mechanics in quantum geometry'',
{\hypersetup{urlcolor=vividviolet}\href{https://inspirehep.net/literature/1737072}{Phys. Lett. B \textbf{794}, 45-49 (2019)}}.

\bibitem{Zhao:2020nrx}
W.~B.~Zhao, G.~R.~Liu and N.~Li,
``Hawking-Page phase transitions of the black holes in a cavity'',
{\hypersetup{urlcolor=vividviolet}\href{https://inspirehep.net/literature/1838422}{Eur. Phys. J. Plus {\bf 136} (2021), 981}}, \href{https://arxiv.org/abs/2012.13921}{[arXiv:gr-qc/2012.13921]}.

\bibitem{Li:2020khm}
  R.~Li and J.~Wang,
  ``Thermodynamics and kinetics of Hawking-Page phase transition'',
{\hypersetup{urlcolor=vividviolet}\href{https://inspirehep.net/literature/1809789}{Phys.\ Rev.\ D {\bf 102}, no. 2, 024085 (2020)}}.


\bibitem{Spallucci:2013jja}
E.~Spallucci and A.~Smailagic,
``Maxwell's equal area law and the Hawking-Page phase transition'',
{\hypersetup{urlcolor=vividviolet}\href{https://inspirehep.net/literature/1257603}{J. Grav. \textbf{2013}, 525696 (2013)}}, \href{https://arxiv.org/abs/1310.2186}{[arXiv:hep-th/1310.2186]}.


\bibitem{Su:2019gby}
B.~Y.~Su, Y.~Y.~Wang and N.~Li,
``The Hawking\textendash{}Page phase transitions in the extended phase space in the Gauss\textendash{}Bonnet gravity'',
{\hypersetup{urlcolor=vividviolet}\href{https://inspirehep.net/literature/1735374}{Eur. Phys. J. C {\bf 80} (2020) no.4, 305}}, \href{https://arxiv.org/abs/1905.07155}{[arXiv:gr-qc/1905.07155]}.

\bibitem{Xu:2020ubo}
Z.~M.~Xu, B.~Wu and W.~L.~Yang,
``Thermodynamics curvature in phase transitions for AdS black hole'',
{\hypersetup{urlcolor=vividviolet}\href{https://inspirehep.net/literature/1814354}{Phys. Lett. B {\bf 821} (2021), 136632}}, \href{https://arxiv.org/abs/2009.00291}{[arXiv:gr-qc/2009.00291]}.


\bibitem{Belhaj:2020mdr}
  A.~Belhaj, A.~El Balali, W.~El Hadri and E.~Torrente-Lujan,
  ``On universal constants of AdS black holes from Hawking-Page phase transition'',
{\hypersetup{urlcolor=vividviolet}\href{https://inspirehep.net/literature/1822951}{Phys.\ Lett.\ B {\bf 811}, 135871 (2020)}}, \href{https://arxiv.org/abs/2010.07837}{[arXiv:hep-th/2010.07837]}.

\bibitem{Wei:2020kra}
  S.~W.~Wei, Y.~X.~Liu and R.~B.~Mann,
  ``Novel dual relation and constant in Hawking-Page phase transitions'',
{\hypersetup{urlcolor=vividviolet}\href{https://inspirehep.net/literature/1802362}{Phys.\ Rev.\ D {\bf 102}, no. 10, 104011 (2020)}}, \href{https://arxiv.org/abs/2006.11503}{[arXiv:gr-qc/2006.11503]}.


\bibitem{Mbarek:2018bau}
  S.~Mbarek and R.~B.~Mann,
  ``Reverse Hawking-Page Phase Transition in de Sitter Black Holes'',
{\hypersetup{urlcolor=vividviolet}\href{https://inspirehep.net/literature/1686333}{JHEP {\bf 1902}, 103 (2019)}}, \href{https://arxiv.org/abs/1808.03349}{[arXiv:hep-th/1808.03349]}.

\bibitem{Marks:2021fpe}
G.~A.~Marks, F.~Simovic and R.~B.~Mann,
``Phase transitions in 4D Gauss\textendash{}Bonnet\textendash{}de Sitter black holes'',
{\hypersetup{urlcolor=vividviolet}\href{https://inspirehep.net/literature/1891560}{Phys. Rev. D \textbf{104}, no.10, 104056 (2021)}}, \href{https://arxiv.org/abs/2107.11352v2}{[arXiv:gr-qc/2107.11352]}.

\bibitem{Nandra:2011ug}
R.~Nandra, A.~N.~Lasenby and M.~P.~Hobson,
``The effect of a massive object on an expanding universe",
{\hypersetup{urlcolor=vividviolet}\href{https://inspirehep.net/literature/897015}{Mon. Not. Roy. Astron. Soc. \textbf{422}, 2931-2944 (2012)}}, \href{https://arxiv.org/abs/1104.4447}{[arXiv:gr-qc/1104.4447]}.


\bibitem{Faraoni:2007es}
V.~Faraoni and A.~Jacques,
``Cosmological expansion and local physics'',
{\hypersetup{urlcolor=vividviolet}\href{https://inspirehep.net/literature/755402}{Phys. Rev. D \textbf{76}, 063510 (2007)
}}, \href{https://arxiv.org/abs/0707.1350}{[arXiv:gr-qc/0707.1350]}.


\bibitem{Hayward:1994bu}
S.~A.~Hayward,
``Gravitational energy in spherical symmetry'',
{\hypersetup{urlcolor=vividviolet}\href{https://inspirehep.net/literature/375366}{Phys. Rev. D \textbf{53}, 1938-1949 (1996)}}, \href{https://arxiv.org/pdf/gr-qc/9408002}{[arXiv:gr-qc/9408002]}.


\bibitem{mass2}
Brien C. Nolan,
``A point mass in an isotropic universe: II. Global properties",
{\hypersetup{urlcolor=vividviolet}\href{https://iopscience.iop.org/article/10.1088/0264-9381/16/4/012}{Class. Quant. Grav. \textbf{16} (1999) 1227}}.

\bibitem{Lake:2011ni}
K.~Lake and M.~Abdelqader,
``More on McVittie's Legacy: A Schwarzschild-de Sitter black and white hole embedded in an asymptotically $\Lambda$CDM cosmology",
{\hypersetup{urlcolor=vividviolet}\href{https://journals.aps.org/prd/abstract/10.1103/PhysRevD.84.044045}{Phys. Rev. D \textbf{84}, 044045 (2011)}}, \href{https://arxiv.org/abs/1106.3666}{[arXiv:gr-qc/1106.3666]}.

\bibitem{Nolan:2014maa}
B.~C.~Nolan,``Particle and photon orbits in McVittie spacetimes",
{\hypersetup{urlcolor=vividviolet}\href{https://inspirehep.net/literature/1309424}{Class. Quant. Grav. \textbf{31}, no.23, 235008 (2014)}}, \href{https://arxiv.org/abs/1408.0044}{[arXiv:gr-qc/1408.0044]}.


\bibitem{Akbar:2006kj}
M.~Akbar and R.~G.~Cai,
``Thermodynamic Behavior of Friedmann Equations at Apparent Horizon of FRW Universe",
{\hypersetup{urlcolor=vividviolet}\href{https://inspirehep.net/literature/726509}{Phys. Rev. D \textbf{75}, 084003 (2007)}}, \href{https://arxiv.org/pdf/hep-th/0609128}{[arXiv:hep-th/0609128]};
M.~Akbar and R.~G.~Cai,
{\hypersetup{urlcolor=vividviolet}\href{https://inspirehep.net/literature/710724}{Phys. Lett. B \textbf{635}, 7-10 (2006)}}, \href{https://arxiv.org/pdf/hep-th/0602156}{[arXiv:hep-th/0602156]}.

\bibitem{Cai:2008ys}
R.~G.~Cai, L.~M.~Cao and Y.~P.~Hu,
``Corrected Entropy-Area Relation and Modified Friedmann Equations'',
{\hypersetup{urlcolor=vividviolet}\href{https://inspirehep.net/literature/674617}{JHEP {\bf 08}, 090 (2008)}}, \href{https://arxiv.org/abs/hep-th/0501055v1}{[arXiv:hep-th/0807.1232]}.


\bibitem{Cai:2008gw}
R.~G.~Cai, L.~M.~Cao, and Y.~P.~Hu,
``Hawking Radiation of Apparent Horizon in a FRW Universe",
{\hypersetup{urlcolor=vividviolet}\href{https://inspirehep.net/literature/796050}{Class. Quant. Grav. {\bf 26}, 155018 (2009)}}, \href{https://arxiv.org/pdf/0809.1554}{[arXiv:hep-th/0809.1554]};
Y.~P.~Hu,
{\hypersetup{urlcolor=vividviolet}\href{https://inspirehep.net/literature/862414}{Phys. Lett. B \textbf{701}, 269-274 (2011)}}, \href{https://arxiv.org/abs/1007.4044}{[arXiv:gr-qc/1007.4044]}.


\bibitem{Saida:2007ru}
H.~Saida, T.~Harada and H.~Maeda,
``Black hole evaporation in an expanding universe'',
{\hypersetup{urlcolor=vividviolet}\href{https://inspirehep.net/literature/751591}{Class. Quant. Grav. \textbf{24}, 4711-4732 (2007)}}, \href{https://arxiv.org/abs/0705.4012}{[arXiv:gr-qc/0705.4012]}.

\bibitem{Gibbons:1977mu}
G.~W.~Gibbons and S.~W.~Hawking,
``Cosmological Event Horizons, Thermodynamics, and Particle Creation'',
{\hypersetup{urlcolor=vividviolet}\href{https://inspirehep.net/literature/125663}{Phys. Rev. D \textbf{15} (1977), 2738-2751}}.

\bibitem{Urano:2009xn}
M.~Urano, A.~Tomimatsu and H.~Saida,
``Mechanical First Law of Black Hole Spacetimes with Cosmological Constant and Its Application to Schwarzschild-de Sitter Spacetime'',
{\hypersetup{urlcolor=vividviolet}\href{https://inspirehep.net/literature/816249}{Class. Quant. Grav. \textbf{26} (2009), 105010}}, \href{https://arxiv.org/abs/0903.4230}{[arXiv:gr-qc/0903.4230]}.


\bibitem{Bhattacharya:2013tq}
S.~Bhattacharya and A.~Lahiri,
``Mass function and particle creation in Schwarzschild-de Sitter spacetime",
{\hypersetup{urlcolor=vividviolet}\href{https://inspirehep.net/literature/1215289}{Eur. Phys. J. C {\bf 73} (2013), 2673}}, \href{https://arxiv.org/abs/1301.4532}{[arXiv:gr-qc/1301.4532]}.


\bibitem{Kubiznak:2015bya}
D.~Kubiznak and F.~Simovic,
``Thermodynamics of horizons: de Sitter black holes and reentrant phase transitions'',
{\hypersetup{urlcolor=vividviolet}\href{https://inspirehep.net/literature/1385749}{Class. Quant. Grav. \textbf{33} (2016) no.24, 245001}}, \href{https://arxiv.org/abs/1507.08630}{[arXiv:hep-th/1507.08630]}.


\bibitem{Cai:2005ra}
R.~G.~Cai and S.~P.~Kim,
``First law of thermodynamics and Friedmann equations of Friedmann-Robertson-Walker universe",
{\hypersetup{urlcolor=vividviolet}\href{https://inspirehep.net/literature/674617}{JHEP \textbf{02}, 050 (2005)}}, \href{https://arxiv.org/abs/hep-th/0501055v1}{[arXiv:hep-th/0501055]}.

\bibitem{Cai:2006rs}
R.~G.~Cai and L.~M.~Cao,
``Unified first law and thermodynamics of apparent horizon in FRW universe",
{\hypersetup{urlcolor=vividviolet}\href{https://inspirehep.net/literature/731742}{Phys. Rev. D \textbf{75}, 064008 (2007)}}, \href{https://arxiv.org/pdf/gr-qc/0611071}{[arXiv:gr-qc/0611071]}.


\bibitem{Gong:2007md}
Y.~Gong and A.~Wang,
``The Friedmann equations and thermodynamics of apparent horizons",
{\hypersetup{urlcolor=vividviolet}\href{https://inspirehep.net/literature/748131}{Phys. Rev. Lett. {\bf 99}, 211301 (2007)}}, \href{https://arxiv.org/abs/0704.0793}{[arXiv:hep-th/0704.0793]}.

\bibitem{Hu:2015xva}
Y.~P.~Hu and H.~Zhang,
``Misner-Sharp Mass and the Unified First Law in Massive Gravity",
{\hypersetup{urlcolor=vividviolet}\href{https://inspirehep.net/literature/1342429}{Phys. Rev. D {\bf 92}, no.2, 024006 (2015)}}, \href{https://arxiv.org/abs/1502.00069}{[arXiv:hep-th/1502.00069]};
R.~G.~Cai, L.~M.~Cao, Y.~P.~Hu, and N.~Ohta,
{\hypersetup{urlcolor=vividviolet}\href{https://inspirehep.net/literature/833807}{Phys. Rev. D \textbf{80} (2009), 104016}}, \href{https://arxiv.org/abs/0910.2387}{[arXiv:hep-th/0910.2387]}.


\bibitem{Ma:2015llh}
M.~S.~Ma and R.~Zhao,
``Stability of black holes based on horizon thermodynamics'',
{\hypersetup{urlcolor=vividviolet}\href{https://inspirehep.net/literature/1402478}{Phys. Lett. B {\bf 751} (2015), 278-283}}, \href{https://arxiv.org/abs/1511.03508}{[arXiv:gr-qc/1511.03508]}.

\bibitem{Hansen:2016ayo}
D.~Hansen, D.~Kubiznak and R.~B.~Mann,
``Universality of $P$-$V$ Criticality in Horizon Thermodynamics'',
{\hypersetup{urlcolor=vividviolet}\href{https://inspirehep.net/literature/1430041}{JHEP {\bf 01} (2017), 047}}, \href{https://arxiv.org/abs/1603.05689}{[arXiv:gr-qc/1603.05689]}.

\bibitem{Kontou:2020bta}
E.~A.~Kontou and K.~Sanders,
``Energy conditions in general relativity and quantum field theory'',
{\hypersetup{urlcolor=vividviolet}\href{https://inspirehep.net/literature/1783879}{Class. Quant. Grav. \textbf{37}, no.19, 193001 (2020)
}}, \href{https://arxiv.org/abs/2003.01815}{[arXiv:gr-qc/2003.01815]}.






\bibitem{Dolan:2013ft}
B.~P.~Dolan, D.~Kastor, D.~Kubiznak, R.~B.~Mann, and J.~Traschen,
``Thermodynamic Volumes and Isoperimetric Inequalities for de Sitter Black Holes'',
{\hypersetup{urlcolor=vividviolet}\href{https://inspirehep.net/literature/1216285}{Phys. Rev. D {\bf 87}, no.10, 104017 (2013)}}, \href{https://arxiv.org/pdf/1301.5926.pdf}{[arXiv:hep-th/1301.5926]}.

\bibitem{Abdusattar:2021wfv}
H.~Abdusattar, S.~B.~Kong, W.~L.~You, H.~Zhang and Y.~P.~Hu,
``Joule-Thomson Expansion and Heat Engine of the FRW Universe",
{\hypersetup{urlcolor=vividviolet}\href{https://inspirehep.net/literature/1909133}{submited to Phys. Lett. B }}, \href{https://arxiv.org/abs/2108.09407}{[arXiv:gr-qc/2108.09407]}.


\bibitem{Kong:2021qiu}
S.~B.~Kong, H.~Abdusattar, Y.~Yin and Y.~P.~Hu,
``The van der Waals-like Phase Transition in the FRW Universe",
{\hypersetup{urlcolor=vividviolet}\href{https://inspirehep.net/literature/1909151}{submitted to JHEP}}, \href{https://arxiv.org/abs/2108.09411}{[arXiv:gr-qc/2108.09411]}.


\end{thebibliography}
\end{document}